\documentclass[12pt]{article}
\title{Foundations of Probability and Physics, Round table.}

\author{Andrei Khrennikov\\
International Center for Mathematical\\
Modeling in Physics and Cognitive Sciences,\\
MSI, University of V\"axj\"o, S-35195, Sweden\\
Email:Andrei.Khrennikov@msi.vxu.se}

\begin{document}
\maketitle

\begin{abstract} We present the summary of the general discussion
on the probabilistic foundations of quantum theory that took place
during the round table at the Int. Conf. "Foundations of Probability 
and Physics", V\"axj\"o, Sweden-2000. It is possible to find at

http://www.msi.vxu.se/aktuellt/konferens/Roundtable.html

continuation of this Round Table. You can send your contribution by  Email
to A. Khr. (subject: Round Table).
\end{abstract}

\section{Interpretations (meanings) of probability}

T. Hida, L. Ballentine, L. Accardi, A. Khrennikov: various probabilistic models can be used to describe various physical phenomena. 
Kolmogorov's model - just one of  possible models for probability. However, technically it is very convenient! 

Change of axiomatics:
T. Hida, L. Ballentine: Þ s-additivity.
L. Accardi: Þ Bayes' axiom for conditional probabilities. 
A. Khrennikov, L. Ballentine: Þ Values of probability, negative, p-adic choice of the "right axiomatics". 
A. Khrennikov: Þ frequency is the only possibility to start. 
L. Ballentine, L. Accardi: Þ not at all!
T. Hida: Þ frequency is not the main element of von Mises theory.

Frequency probability, Laplace probability, ensemble probability, subjective, …
Different models. Why do they give the same answer?

L. Ballentine: I am surprised.
A. Khrennikov: It is not always: for example,  p-adic frequency and ensemble probabilities do not coincide.
T. Hida: new mathematics, new probability Þ  new physics!
V. Maximov: randomness is the basic concept that generates probability.
L. Accardi: pay attention on the Russian variant of Kolmogorov's book: independence and non-measurable sets.
A. Khrennikov: connections with Gudder's theory?
S. Gudder: Not clear…
A. Khrennikov: frequency and ensemble probabilities coincide due to ergodicity. It may be that ergodicity hypothesis is violated in quantum mechanics? Bounomano?
J. Summhammer: It seems not!

\section{Interpretations (meanings) of wave function}

W. De Muynck: empirisists ("contextualist") interpretation is the "best one". No problems, no paradoxes. In particular, no nonlocality. 
L. Accardi: "camelion model" (a kind of local contextualism) 
P. Lahti: "Well contextualism is okey". But we can try to find some deeper level of interpretation. Operator valued measures. Individual interpretation? Experimental physics works with individual systems. 
L. Ballentine: interpretation of wave function ~ interpretation of probability.
L. Ballentine: started with the realist ensemble interpretation Þ contextualist ensemble interpretation. However, realist ensemble interpretation is also possible (depending on probabilistic model).
J. Larsson: Bohmian pilot wave interpretation.
A. Kracklauer: pilot wave (as De Broglie?)
S. Aerts: ?
B. Coecke: ?

\section{ Mystery of quantum mechanics}

W. De Muynck, L. Accardi, L. Ballentine: no mystery, everything depend on contexts of preparations and measurement. All probabilities are conditional. This explain the violation of the classical probabilistic rule. 
I. Volovich: Well explain violations. But, please, explain the origin of quantum interference. Two slit experiment. 

Long discussion.

J. Summhammer: On interference of classical balls.
L. Accardi, J. Summhemmer: explanations via quantum theory.
I. Volovich, A. Khrennikov: do not  like all these explanations. We need some explanation on sub-quantum level.
A. Khrennikov: quantum (and more general, in particular, "hyperbolic") probabilistic rule via classical frequency model.  However, the only physical explanation is based on perturbation effects of preparation procedures. Again contextualism.
A. Khrennikov: Contextualism Þ nonlocality in the two slit experiment.
A. De Muynck, Accardi, Ballentine, Summhammer: against this.
A. Khrennikov: It is not clear, how we can explain that by closing one slit we can change behaviour of localized particle near another slit?
W. De Muynck: It might in principle be that there is something. For example, by closing slit N1 we could change vacuum fluctuations near slit N2. 
J. Summhammer: It would be impossible. Quantum mechanical explanation via Schrödinger equation. 
A. Kracklauer: pilot wave explanation of interference. 
I. Volovich: no real space-time analysis in Bell's experiment. New mathematics, for example, p-adic New models of space-time, for example, p-adic space. New viewpoints to locality. 
A. Khrennikov: p-adic probability Þ p-adic elements of reality. Our modern picture of physical reality is based on real numbers: real space Þ  real locality, real statistical stabilization ('law of large numbers') Þ real elements of reality. 

Long discussion on mystery of quantum mechanics, more and more chaotic, …, total chaos, …, the end of the discussion.

\section{Post-table remarks}

1. Prof. A. F. Kracklauer : Probability in Quantum Mechanics
The Born interpretation of the wave function as a entity whose intensity is
a probability density evokes two issues:
Reconciling the conflict between the incompleteness implicit in all definitions
of a probability with the generally accepted view that QM itself is complete.
Reconciling the appearance of interference of wave functions with the fact that
all customary applications of probability theory do not exploit this structure.
Following are this writer's personal attitude to these issues as refined by
his participation in the V\"axj\"o Conference. They constitute the sort of remarks
that he wishes in hindsight he had made while participating. In some cases they
actually agree with remarks he did make.
Completeness: Problem definition:

All definitions of probability in the end reduce to the ratio of occurrence
of events from a restricted set to the unrestricted set. To the extent that
this definition is not optimal for developing abstract structure, some researchers,
mostly mathematicians, have crystallised certain properties of probabilities
to use as primary elements in developing probability theory; i.e., proving theorems.
Perhaps Kolmogorov is the most renown example. Nevertheless, whenever applying
results so developed, the underlying concept is inevitably this ratio. Most,
perhaps all, statements about probabilities can be rendered in terms of this
ratio.

For example, the admonition not to forget that all probabilities are conditional
probabilities is essentially the statement that in applications to a particular
problem, the denominator of the ratio must be correctly identified. In other
words, within a problem one must continuously verify that the unrestricted set
is the same. Likewise, the various definitions of probability can be seen as
recipes for determining the numerator. The ``ensemble'' approach essentially
requires materially counting the members of the restricted set (at least conceptually).
``Geometric'' probability uses some knowledge of the events to \emph{calculate}
the size of the numerator (useful for physics applications where the sets are
so large that mortals will never count them). For the `propensity' approach,
the numerator is simply intuited, either by the calculator or perhaps by Mother
Nature herself. 

All these approaches in the end imply some sort of incomplete knowledge---at
least to the extent that probabilities with values other that \( 0 \) or \( +1 \)
are involved, and QM with no more seems unimaginable. While the `propensity'
approach seems to offer a reconciliation with the completion problem by putting
the task of determining the numerator on a spooky, unidentified agent or process,
any attempt to quantify `propensity' ultimately reduces to the frequency definition
or vanishes into a semantic swamp.

Possible resolution:
In this writer's view, the Stochastic Electrodynamics (SED) model of QM offers
the optimum resolution of the problem by identifying the wave function as an
extended sort of pilot wave. The SED pilot wave differers from its predecessor
as introduced by De Broglie in that it is modulation on \emph{Zitterbewegung}
which is in turn evoked by a stochastic classical electromagnetic background.
In this paradigm, it is argued that relative motion through the background evokes
a modulation at the scale of De Broglie waves on a Brownian \emph{zitter}motion
on the scale of Compton wavelengths. When this effect is taken into account
for ensembles of similar systems, the Liouville Equation for the ensemble is
so modified that it implies the Schr\"odinger Equation. In the end, it is seen
that wave functions intensities are related to Liouville densities so that there
is no fundamentally new probabilistic notion introduced by QM. In effect, the
Einstein, Podolsky and Rosen surmise that QM (at the De Broglie level vice the
\emph{Zitter} level) is incomplete is confirmed, thereby preempting conflicts
with the definition of probability.
Interference in Probability:

Numerous analysts from various view points have (re)discovered that fact that
Probability Theory admits structure (used by QM) that goes unexploited in traditional
applications. This fact is among the most oft reinvented wheels in all of Physics;
examples known to and easily recalled by this writer include:

from an instrumentalist approach: Kershaw, Collins, Nelson;
from probability theory: Gudder, Collins;
from an empiricist point of view: Summhammer.

While each of these approaches provides deep and surprising insights, none really
offers any explanation of why and how nature is exploiting this structure. Just
as a certain second order hyperbolic partial differential equation becomes \emph{the}
``wave equation,'' as a physics theory only with the introduction; e.g., of
Hook's Law, so this extra probability structure can be made into physics only
with a analogue to Hook's Law. 

SED provides that analogue for particle behaviour with its model of pilot wave
guidance. In this model, radiation pressure is responsible for particle guidance.
Radiation pressure is proportional to the \emph{square} of EM fields; i.e.,
the intensity (in this case of the the background field as modified by objects
in the environment) which is not additive. Rather the field amplitudes are additive
and interference arrises in the way well understood in classical EM. In other
words, QM interference is a manifestation of EM interference. The relevant Hook's
Law analogue is the phenomenon of radiation pressure. For radiation, this is
all intimately related, of course, to classical coherence theory as applied
to square law'' photoelectron detectors, which, when properly applied,
resolves many QM conundrums, including those instigated by Bell's Theorem surrounding
EPR correlations.

2. V.I.Serdobolskii:   ON PHYSICS OF QUANTUM  PROBABILITY

      Quantum mechanics is the wave mechanics and its evolution
equations are not much different from those for electromagnetic
field or acoustics. The wave function is the field amplitude being
complex as in electrodynamics, and the square of its modulus
presents the concentration of matter.
There are no problems
of the sort: "how could an electron feel the second slit in
two-slit experiments": it is the well known interference effect.

The experiments of neutron scattering on a nucleus
are well described in so called optical model, in which the
nucleus is considered as a half-transparent body.
However, the optical scattering is the scattering of beams.
But we can observe the scattering of separate neutrons.
If the De Broglie wave length of the neutron is small,
then it is well described by geometrical
optics. If the wavelength is large, then experiments show typical
pictures of a diffraction. In the scattering of beams of particles,
the diffraction picture can be interpreted as a statistical picture
of a summation of elementary scattering processes, and quantum mechanics
works as a statistical theory.
But in the scattering of separate neutrons,
the diffraction picture only presents a probability distribution.
These separate observations cannot be predicted by quantum mechanics.
The wave equations yield only "waves of probability."
These features are most specific for a new phenomenon
which we would call "quantum probability".

And here we meet a fundamental problem: is the observed trace of a
particle principally unpredictable and probabilistic,
or it can be predicted deterministically in some unknown extension
of physical theories?

Physicist knows well the relative value of theories.
After a while, unpredictable phenomena prove to be explained
and well described in other more refined theories.
And a serious investigator would seek an explanation
of the behaviour of separate particles first in a deeper insight
into physics of the micro world.

One could seek the causality in the effect of initial conditions
which are hardly measurable for elementary particles.
However, there are quantum effects when initial conditions
are well known (for example, in optics), but the behaviour of
separate particles cannot be predicted.

The constructive approach to probability theory shows that the
there must exist a complicated process of generating
randomness that we observe as the quantum probability.
And it is quite natural to search it in the dynamics
of inner state of particles.
Most probably, particles present rather complicated systems
"living their own lives". 
Let us recall that in modern quantum field theory,
each micro particle is viewed as a source of a cloud of different 
virtual particles that are permanently born and annihilated.
So called elementary particles, most probably, present complicated 
evolutioneering quantum dynamical systems that we observe today
by measuring only a small number of parameters.

\end{document}